\documentclass[%
]{algotel}

\usepackage{wrapfig}
\usepackage{tikz}
\usepackage{siunitx}

\usetikzlibrary{arrows,
                calc,
                positioning,
                quotes}
\usetikzlibrary{shapes.misc}
\usetikzlibrary{shapes.geometric}
\usetikzlibrary{patterns}
\usetikzlibrary{decorations.pathreplacing}

\author{Quentin Bramas\addressmark{1}
  \and Jean-Romain Luttringer\addressmark{1}
  \and Pascal Mérindol\addressmark{1}}

\title{La ROUTOURNE va tourner : le retour des tours de détours sous leurs plus beaux atours}

\address{\addressmark{1} Université de Strasbourg, Laboratoire ICube}


\def\preceqdot{\mathrel{\prec\kern-.5em\raise.05ex\hbox{$\cdot$}}} 
\def\lhddot{\mathrel{\lhd\kern-.35em\raise.035ex\hbox{$\cdot$}}} 
\def\lhdeq{\mathrel{\leqslant\kern-.498em\raise.255ex\hbox{$\shortmid$}}}

\usepackage[T1]{fontenc}
\usepackage{xspace}
\usepackage{hyperref}
\newcommand{\routourne}{{R%
\kern-.15em\raise0.3ex\hbox{\small O}%
U%
\kern-.15em\raise0.3ex\hbox{T}%
\kern-.15em\hbox{O}%
\kern-.15em\raise0.3ex\hbox{\small U}%
\kern-.05em\hbox{R}%
\kern-.1em\lower0.2ex\hbox{\small N}%
\kern-.08em\raise0.38ex\hbox{\small E}}\xspace}
\begin{document}
\maketitle

\begin{abstract}

Le routage par segments (SR) offre un contrôle précis sur les chemins empruntés : il spécifie dans les paquets IP une liste de détours, appelés \textit{segments}. Le nombre de détours pouvant être spécifiés est cependant limité par le matériel. Lors du calcul de listes de segments, il est donc nécessaire d'en limiter la taille. Bien que des solutions aient été proposées pour calculer ces listes, celles-ci manquent de généralité, et ne sont pas toujours optimales ou efficaces. 
Nous présentons \routourne, une méthode permettant de détourner les algorithmes de routage afin que ces derniers calculent, non pas simplement un chemin physique optimal à traduire en liste de segments a posteriori (sans garantie sur sa taille), mais directement les listes de segments optimales et déployables par le matériel sous-jacent. \routourne facilite ainsi le déploiement de stratégies d'ingénierie de trafic et de politiques avancées, notamment pour équilibrer la charge depuis les sources. 
Malgré une route truffée de défis surprenants -- en particulier, la perte d'isotonie induite par SR, \routourne s'avère efficace, induisant au pire un surcoût linéaire. Son exactitude et optimalité ont été prouvées, et son efficacité évaluée en la généralisant à plusieurs algorithmes de calcul de chemins plus ou moins complexes.

\end{abstract}

\vspace{-0.3cm}
\section{Intourduction}\label{sec:intro}

Les réseaux IP utilisent le routage "best-effort" au saut par saut, où les chemins minimisent le \emph{coût IGP}, une métrique additive définie par l'opérateur. Cependant, il est fréquent de devoir s'affranchir de ces chemins pour leur privilégier ceux offrant des garanties concrètes, comme une latence bornée pour certains flux premium ou le contournement d'équipements défaillants.

Le calcul de tels chemins est un défi en soi, surtout lorsque plusieurs métriques sont prises en compte simultanément. Cependant, nous nous penchons ici sur l'étape suivante, souvent négligée : le déploiement de ces chemins contraints en détournant le trafic des chemins "best-effort" utilisés dans le réseau.


En pratique, ces détours sont construits à l'aide du routage à la source relâché, dont l'implémentation la plus répandue est Segment Routing (SR). SR permet de spécifier directement dans le paquet une liste de détours. Le paquet va de détour en détour en suivant les plus courts chemins IGP entre ces derniers. Cependant, le nombre de détours (appelés "segments") spécifiables est limité par le matériel sous-jacent. Cette contrainte, appelée Profondeur Maximale des Segments (PMS), implique que certains chemins physiques peuvent ne pas être déployables (car nécessitant trop de segments pour être décrits). 
Cependant, prendre en compte cette contrainte lors du calcul des chemins pose plusieurs défis 
qui ont des conséquences inattendues, comme la perte d'optimalité des sous-chemins (ou isotonie), et rendent impossible l'utilisation naïve des algorithmes existants. Il est donc nécessaire de fournir un cadre correct et efficace pour intégrer SR sans compromettre les performances des calculs.

Nous présentons \routourne, une méthode facilitant le déploiement d'ingénierie de trafic avancée. \routourne permet d'adapter des algorithmes de calculs de chemins existants afin que ces derniers retournent des listes de segments optimales et déployables. \routourne repose sur deux éléments clés :

\begin{itemize}
    \item La traduction des chemins explorés en listes de segments minimales \emph{pendant l'exploration du graphe afin de guider la recherche}; 
    \item La modification de la fonction de comparaison des chemins afin de considérer le nombre de segments correctement, malgré la perte de l'isotonie standard.
\end{itemize}

L'optimalité et l'exactitude de \routourne ont été démontrées et son efficacité évaluée sur des réseaux variés, exhibant son faible surcoût linéaire. Le code de \routourne et des expériences est public~\footnote{\url{https://zenodo.org/records/10202365}}.

\section{Défis et tour de l'état de l'art}
\begin{wrapfigure}{l}{0.59\textwidth} 
    \centering
       
\begin{tikzpicture}[yscale=1.3, xscale=1.3, -,>=stealth',shorten >=1pt,auto,node distance=3cm,
    thick,main node/.style={circle,fill=white,draw,font=\sffamily,inner sep=3pt},
    edgelabel/.style={fill=white,inner sep=1pt, anchor=center, pos=0.5,font=\sffamily}]

\begin{scope}[xshift=-1cm, yshift=0.1cm]\footnotesize
\node[main node, fill=black!20] (S) at (-0.5,1.7) {S};
\node[main node] (1)                at (0.3,0.8) {1};
\node[main node,line width=1pt, draw=orange] (6)                at (1.2,0) {4};
\node[circle,line width=1pt,draw,font=\sffamily,inner sep=3pt, draw=black!20!green,dashed] (6)                at (1.2,0) {4};
\node[main node,draw=blue] (2)                at (1,1.7) {2};
\node[main node,fill=black!20] (3)  at (2,0.8) {3};
\node[main node,fill=black!20] (D)  at (4.5,0.8) {D};
\node[font=\sffamily] at (-0.47,1.4) {\scriptsize a};
\node[font=\sffamily] at (-0.28,1.5) {\scriptsize b};

\draw (S) [->, bend left=20] edge[] node[edgelabel] {1,2} (1);
\draw (S) [->, bend left=-20] edge[] node[edgelabel] {1,1} (1);

\draw (1) [->] edge[] node[edgelabel] {1,1} (6);
\draw (1) [->] edge[] node[edgelabel] {1,4} (3);
\draw (6) [->] edge[] node[edgelabel] {1,1} (3);

\draw (3) [->] edge[] node[edgelabel] {2,1} (D);

\draw (S) [->] edge[] node[edgelabel] {2,1} (2);
\draw (2) [->] edge[] node[edgelabel] {1,2} (3);
\draw (2) [->, bend left=20] edge[] node[edgelabel] {2,6} (D);

\end{scope}

\footnotesize
\begin{scope}[xshift=-0.8cm]
\node[rectangle, draw=blue] (d1) at (2.3,0.4) {$3,\textbf{3}\,(2)$};
\node[rectangle, draw=black!20!green] (d2) [below=0.55 of d1.west,anchor=west] {$3,\textbf{3}\,(3)$};
\node[rectangle, draw=orange] (d3) [below=0.55 of d2.west,anchor=west] {$3,\textbf{4}\, (2)$};
\node[rectangle, draw] (d4) [below=0.55 of d3.west,anchor=west] {$2,\textbf{6}\, (1)$};
\end{scope}
\node[] (d1-d) [right=0.1 of d1]  {2-3};
\node[] (d2-d) [below=0.55 of d1-d.west,anchor=west] 
{$1_a$-4-3};
\node[] (d3-d) [below=0.55 of d2-d.west,anchor=west] 
{4-3};
\node[] (d4-d) [below=0.55 of d3-d.west,anchor=west] 
{3};

\begin{scope}[xshift=0.7cm]
\node[rectangle, draw=blue] (d11) at (3.2,0.4) {$5,\textbf{4}\, (3)$};
\node[rectangle, draw=black!20!green] (d12) [below=0.55 of d11.west,anchor=west] 
{$5,\textbf{4}\, (3)$};
\node[rectangle, draw=orange] (d13) [below=0.55 of d12.west,anchor=west] 
{$5,\textbf{5}\, (2)$};
\node[rectangle, draw] (d14) [below=0.55 of d13.west,anchor=west] 
{$4,\textbf{7}\, (1)$};
\end{scope}

\node[] (d21) [right=0.1 of d11]  {2-3-D};
\node[] (d22) [below=0.55 of d21.west,anchor=west] 
{$1_a$-4-D};
\node[] (d23) [below=0.55 of d22.west,anchor=west] 
{4-D};
\node[] (d24) [below=0.55 of d23.west,anchor=west] 
{D};

\draw (-1.5,-0.5) edge[->] node[edgelabel] {\scriptsize IGP, Latence} (0.2,-0.5);
\node[rectangle, draw] at (-0.6,-0.8) {\scriptsize IGP, \textbf{Latence} (\#seg)};

\end{tikzpicture}
    \vspace{-4mm}
    \caption{Segments nécessaires pour encoder les distances optimales vers 3 et D. \routourne résout le défi de l'isotonie perdue.}
   \vspace{-5mm}
\label{fig:iso}
\end{wrapfigure}
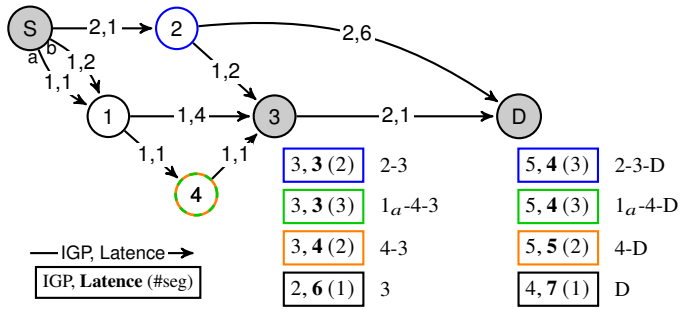

L'approche type pour obtenir une ou des listes de segments est d'\textit{encoder} un unique chemin donné en entrée en un nombre minimal de segments~\cite{7778603}. Bien qu'utile pour certains scénarios d'usage, ce modèle de traduction a posteriori présente des limites.
Il ne tire pas parti de l'opportunité d'équilibrer la charge sur des chemins équivalents (e.g., offrant la même latence) et, plus problématique, il est dissocié du calcul du chemin. Il n'y a donc aucune garantie que le chemin passé en entrée soit encodable en moins de PMS segments, comme l'illustre la figure~\ref{fig:iso}. Encoder le chemin de plus faible latence de S vers D résulte en une liste de segments de taille 3, qui pourrait ne pas être déployable sur du matériel fortement contraint. De ces observations émerge le premier défi : \textbf{intégrer directement l'algorithme d'encodage dans le calcul de chemins afin d'en guider la recherche en tirant partie de l'équilibrage de charge.}

Supposons celui-ci résolu, avec la connaissance du surcoût opérationnel permettant de déployer les distances explorées. Plusieurs métriques sont ainsi à considérer : non seulement les métriques du problème originel, mais aussi le nombre de segments nécessaires. Le calcul de telles distances multi-métriques partage le même principe que son homologue mono-métrique : chaque distance optimale est étendue afin de déterminer les meilleures distances vers des noeuds plus éloignés. La définition d'\emph{isotonie} est cependant généralisée afin d'inclure l'ensemble des distances non-dominées, i.e., le \emph{front de Pareto} des distances\footnote{Distances offrant un compromis intéressant, pour lesquelles il n'existe aucune distance meilleure sur l'intégralité des métriques.}. Intuitivement, modifier le calcul afin d'étendre toutes les distances non-dominées (nombre de segments compris) semble suffisant pour garantir le calcul correct des listes de segments déployables. 

Pourtant, cela ne garantit pas de découvrir toutes les solutions optimales, tel qu'illustré en Figure~\ref{fig:iso}. Supposons que l'on recherche le chemin de plus faible latence. Plusieurs distances mènent au nœud 3. La distance en bleue $(3,3 (2))$ domine les distances verte et orange ; ces dernières possédant des poids égaux ou pires. Selon le principe d'optimalité des sous-chemins, seule la distance bleue serait alors étendue. Cependant, une fois la distance bleue étendue, un segment supplémentaire est nécessaire pour encoder celle-ci $(5,4 (3))$ jusqu'au noeud D. En effet, le détour $3$ doit être explicitement spécifié, au risque de laisser un paquet utiliser le lien $(2, D)$, dont la latence proscrit l'utilisation pour garantir la distance $(5,4)$. En revanche, les distances orange et verte peuvent être étendues jusqu'à D sans nécessiter de segment supplémentaire. \emph{Repousser} leur dernier détour (nœud $3$) d'un saut (jusqu'à D) est suffisant pour garantir que les chemins utilisés possèdent au pire une distance de $(5,4)$. Ainsi, la distance verte, originellement dominée par la distance bleue, devient une solution de qualité égale. Plus intéressant encore, la distance orange devient la seule distance encodable en 2 segments offrant une latence proche de l'optimale. Cette distance, pouvant être la seule déployable sur du matériel contraint, ne sera trouvée qu'en étendant une distance sous-optimale dominée.

Cette perte d'isotonie pose problème : si l'ensemble des distances non-optimales devaient  être étendues, la complexité du calcul pourrait croître exponentiellement. Certaines méthodes de calcul de listes de segments contournent ce problème en utilisant un graphe transformé où la métrique du nombre de segments est isotone, ce qui permet l'utilisation de méthodes de calcul de chemins classiques~\cite{AubryPhD, LUTTRINGER2022109015, Lazzeri2015EfficientLE}. Cependant, le graphe transformé est une clique, et cette densité maximale génère un surcoût non négligeable pour la plupart des algorithmes.
Il en résulte un second défi : \textbf{garantir que les meilleures listes de segments déployables seront correctement et efficacement calculées.}


\section{ROUTOURNE : encoder et comparer à tour de bras}

Afin de calculer des listes de segments optimales déployables, \routourne repose sur deux ingrédients : un algorithme d'encodage intégré et efficace ainsi qu'une modification de la comparaison des chemins afin de regagner l'isotonie perdue. Pour rester concis, nous présentons \routourne informellement en simplifiant certains principes. L'intégralité du cadre formel est disponible dans le rapport technique~\cite{bramas2023simple}.

\paragraph{Encoder les distances.} Nous proposons un algorithme glouton pour l'encodage des distances à la volée. Celui-ci vérifie, selon les besoins, les propriétés spécifiées comme par exemple contourner un élément en panne. \`A chaque extension de distance, notre algorithme évalue si la nouvelle distance découverte est encodable en un unique segment (i.e., en repoussant le dernier détour spécifié d'un saut supplémentaire). Il est alors nécessaire de s'assurer que les chemins sous-jacents au segment utilisé vérifient les propriétés désirées, et possèdent des distances égales ou meilleures à la distance considérée. Dès lors que l'extension par une nouvelle arête rend la distance non-encodable par un unique segment, un segment (i.e., détour) supplémentaire est ajouté. Lors de la prochaine extension, l'algorithme d'encodage suit le même principe, mais considère ce détour intermédiaire comme nouvelle source. Une organisation efficace de la base de données des segments permet de réaliser cette vérification pour un coût négligeable. 

Bien que l'encodage soit guidé par le chemin en cours d'exploration par l'algorithme, notre méthode autorise des \textit{écarts} si ces derniers satisfont les propriétés et distances du chemin considéré. Ainsi notre encodage permet de tirer parti de l'équilibrage de charge sur des chemins au moins aussi \textit{bons que le guide}.

\paragraph{Redéfinition de la non-dominance.} Comme expliqué et illustré dans les parties précédentes, étendre uniquement les distances non-dominées est insuffisant pour garantir la découverte de toutes les distances optimales avec SR. Considérer l'ensemble des distances comme potentiellement optimales et toutes les étendre n'est pas une option réaliste en terme d'efficacité. Dans \routourne, nous révisons la définition de l'ensemble des distances à étendre pour assurer la découverte efficace des solutions optimales.

Les métriques standards telles que le coût IGP ou la latence restent isotones, seul le nombre de segments se comporte comme une métrique inhabituelle. \routourne analyse donc les cas dans lesquels des distances dominées pourraient nécessiter un nombre de segments égal à, ou meilleur que leurs concurrentes après extension, i.e., qui ne requerront pas de segment supplémentaire une fois étendues.

Le première observation essentielle est que le motif décrit ci-dessus, dans lequel l'écart en nombre de segments nécessaires pour une distance est réduit d'un segment par rapport à ses concurrentes, ne peut pas se produire consécutivement~\footnote{La preuve se déduit de l'optimalité des sous-chemins. Intuitivement, une listes de segments $L'$ ayant été récemment augmenté d'un segment explore des chemins inclus dans les chemins explorés par une liste dont le détour est en amont. Par inclusion, si $L'$ requiert un segment supplémentaire pour garantir une certaine distance, $L$ également.}. Une distance dominée nécessitant deux segments (ou plus) que ses concurrentes ne pourra donc jamais devenir optimale, et peut être supprimée de l'exploration sans risque.
 
La seconde observation est que des listes de segments partageant le même dernier détour évolueront de la même manière. En effet, ces listes construiront leurs détours successifs à partir des plus courts chemins IGP depuis une source intermédiaire commune. Par conséquent, il n'y a aucune raison qu'une de ces listes en particulier requière un segment supplémentaire et pas les autres. Ainsi, les distances dont les listes de segments partagent le même dernier détour peuvent être comparées selon les règles de dominance habituelles.


Ces deux observations nous permettent de conclure qu'une distance sous-optimale peut devenir optimale après extension uniquement si elle nécessite au pire un segment de plus que les distances qui la dominent, et ce alors que son dernier segment a une source différente de celles des listes qui la dominent. Seules ces distances additionnelles doivent être étendues, même si elles sont dominées au sens usuel du terme.

Cette relation de dominance \emph{étendue} est à nouveau isotone, mais augmente le nombre de distances à considérer d'un facteur $|V|$ au pire, $V$ étant l'ensemble des noeuds du graphe (au pire, une distance dominée supplémentaire par source intermédiaire doit être maintenue). En pratique, ce surcoût dépasse rarement un facteur de 3 sur des graphes réalistes. Enfin, cette relation peut être ajustée selon plusieurs modèles pour obtenir une diversité maximale de listes de segments si l'opérateur souhaite bénéficier d'un large équilibrage de charge à la source (notre rapport technique~\cite{bramas2023simple} donne davantage d'exemples).

\paragraph{Détourner l'algorithme} En résumé, \routourne modifie l'algorithme originel afin de calculer efficacement des listes de segments optimales et déployables car (i) les distances explorées sont directement encodées en listes de segment à chaque extension (et non a posteriori), et (ii) seules les distances additionnelles vérifiant notre relation de dominance étendue sont maintenues et actualisées.

\section{\'Evaluation : ROUTOURNE tourne plutôt bien}


\begin{figure}
  \centering
  \begin{minipage}[b]{0.49\textwidth}
    \centering
    \includegraphics[scale=0.39]{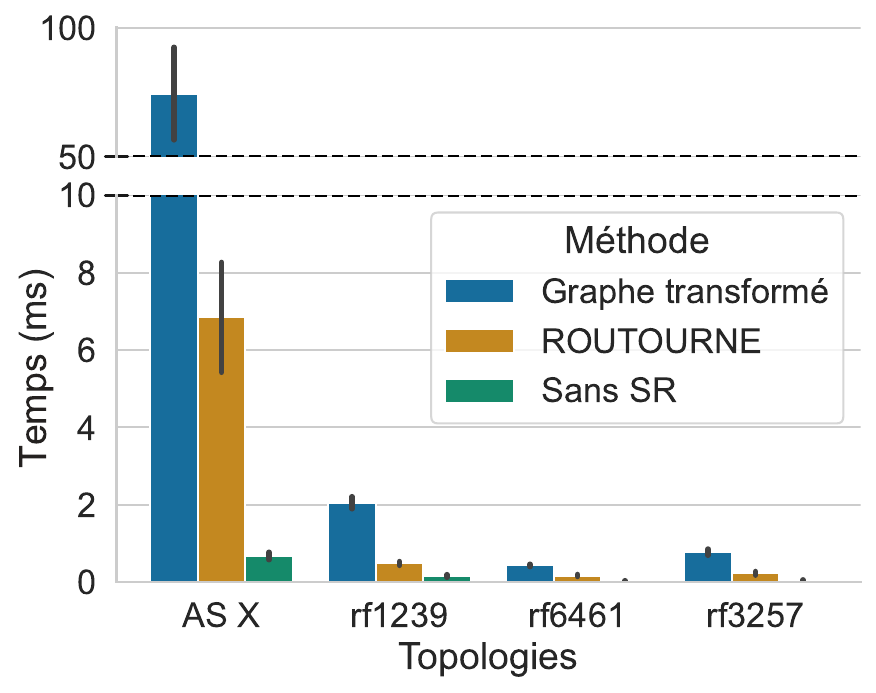}
    \vspace{-0.2cm}
    \caption{\'Evaluation sur topologies réalistes de 2000, 1000, 370 et 300 arêtes respectivement.}
    \label{fig:sub1}
  \end{minipage}
  \hfill
  \begin{minipage}[b]{0.49\textwidth}
    \centering
    \includegraphics[scale=0.39]{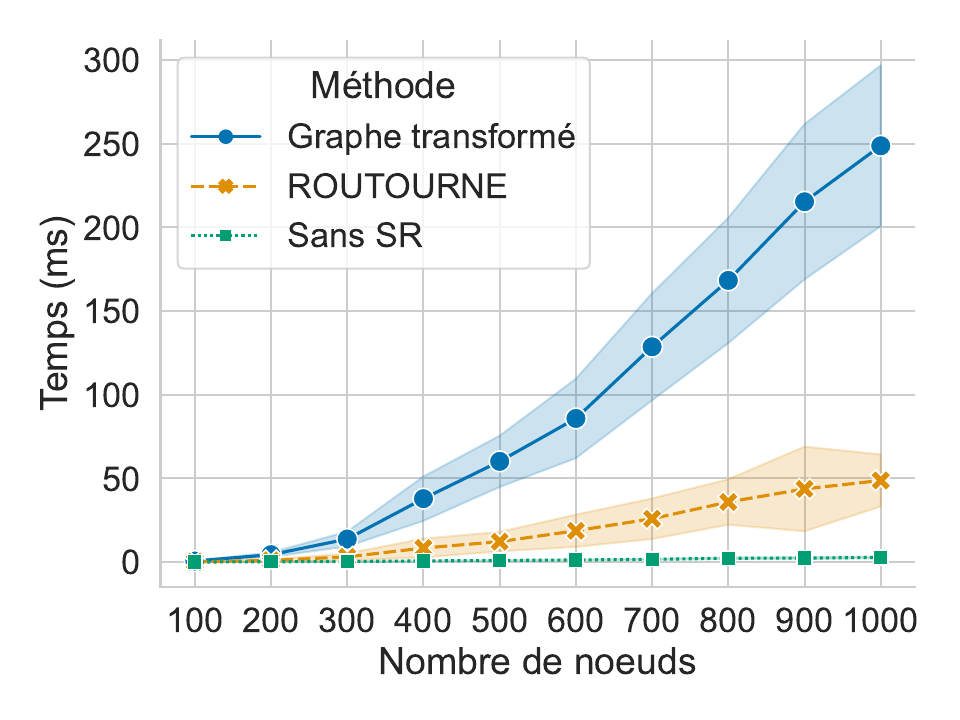}
    \vspace{-0.3cm}
    \caption{\'Evaluation du temps de calcul sur des graphes synthétiques et denses.}
    \label{fig:sub2}
  \end{minipage}
\end{figure}

Les garanties théoriques de \routourne étant formellement démontrées~\cite{bramas2023simple}, cette évaluation se concentre sur la question de l'efficacité : \emph{Quel est l'impact des modifications apportées par ROUTOURNE sur les performances de l'algorithme originel ?} Nous adaptons un algorithme multi-métrique, SAMCRA, de deux manières différentes : en utilisant \routourne et en lançant SAMCRA sur un graphe transformé (une clique), où le nombre de segments est nativement isotone. Notre analyse se concentre sur un cas difficile : le calcul des plus court chemins IGP respectant une borne supérieure sur la latence. Nous menons cette étude sur des graphes réalistes et synthétiques denses, afin de dégager d'éventuelles tendances. 

Les résultats, présentés dans les figures \ref{fig:sub1} et \ref{fig:sub2}, montrent que bien que \routourne augmente le temps d'exécution de l'algorithme (ce qui est attendu, puisque le problème devient multi-métrique), il reste nettement plus efficace que l'approche basée sur la transformation de graphe, notamment pour les réseaux de grande taille. Dans certains cas, le surcoût introduit par \routourne semble même négligeable.

\section{Pour clôtourer}

\routourne est une méthode générique augmentant les algorithmes de calcul de chemins afin qu'ils retournent des listes de segments optimales et déployables, facilitant ainsi l'ingénierie de trafic avancé. Elle repose sur deux éléments : un algorithme d'encodage efficace et relâché pour traduire les chemins en listes de segments, et une modification de la fonction de comparaison des distances. Cette redéfinition garantie la découverte efficace de solutions optimales malgré la perte d'optimalité des sous-chemins, fournissant diverses listes de segments immédiatement déployables selon l'équilibrage de charge désiré.

{
\scriptsize
\bibliographystyle{alpha}
\vspace{-1mm}
\bibliography{ref}
\label{sec:biblio}
}
\end{document}